\documentclass[preprint,12pt]{elsarticle}

\usepackage{graphicx}
\usepackage{amssymb}
\usepackage{hyperref}
\hypersetup{colorlinks,
linkcolor=blue,
filecolor=green,
urlcolor=blue,
citecolor=blue}

\journal{Annals of Physics}
\begin{document}

\begin{frontmatter}

\title{A microscopic approach to Casimir and Casimir-Polder forces between metallic bodies}

\author{Pablo Barcellona\fnref{presentaddress}}

\fntext[presentaddress]{Present address: Institute of Physics, Albert-Ludwigs University of Freiburg, Hermann-Herder-Str. 3, D-79104 Freiburg, Germany.
E-mail address: pablo.barcellona@physik.uni-freiburg.de}

\author{Roberto Passante\fnref{email}}
\address{Dipartimento di Fisica e Chimica, Universit\`a degli Studi di Palermo and CNISM, Via Archirafi 36, I-90123, Palermo, Italy}
\fntext[email]{E-mail address: roberto.passante@unipa.it}

\begin{abstract}
We consider the Casimir-Polder interaction energy between a metallic nanoparticle and a metallic plate, as well as the Casimir interaction energy between two macroscopic metal plates, in terms of the many-body dispersion interactions between their constituents. Expressions for  two- and three-body dispersion interactions between the microscopic parts of a real metal are first obtained, both in the retarded and non-retarded limits. These expressions are then used to evaluate the overall two- and three-body contributions to the macroscopic Casimir-Polder and Casimir force, and to compare them with each other, for the two following geometries: metal nanoparticle/half-space and half-space/half-space, where all the materials are assumed perfect conductors. The above evaluation is obtained by summing up the contributions from the microscopic constituents of the bodies (metal nanoparticles). In the case of nanoparticle/half-space, our results fully agree with those that can be extracted from the corresponding macroscopic results, and explicitly show the non-applicability of the pairwise approximation for the geometry considered. In both cases, we find that, while the overall two-body contribution yields an attractive force, the overall three-body contribution is repulsive.  Also, they turn out to be of the same order, consistently with the known non applicability of the pairwise approximation. The issue of the rapidity of convergence of the many-body expansion is also briefly discussed.
\end{abstract}


\end{frontmatter}


\section{Introduction}
Casimir and Casimir-Polder forces are electromagnetic interactions between neutral macroscopic bodies or between atoms and macroscopic bodies, respectively, due to the quantum zero-point fluctuations of the electromagnetic field \cite{Casimir48,Milonni94}.
Dispersion forces are analogous interactions between neutral microscopic polarizable objects, such as atoms, molecules, condensates or nanoparticles, and described in terms of exchange of virtual photons between them or in terms of vacuum fluctuations \cite{CP48,Buhmann,Buhmann2,CPP95}. Dispersion forces (van der Waals, including the Casimir-Polder regime) are not additive \cite{AT43,AZ60,PT85,CP97,RRE09,Salam10,MA14}, and many-body effects may become relevant for dense systems \cite{PP99,SS12}.

It is worth to consider Casimir and Casimir-Polder forces involving macroscopic bodies in terms of the dispersion interactions between their constituent parts and investigate the role of non-additive components. This could also give a deeper understanding on the origin of Casimir forces, showing how they derive from microscopic fundamental interactions.
For example, in the macroscopic approach the presence of matter interacting with a quantum field is usually taken into account by the boundary conditions it imposes on the field operators.
Some models aiming to obtain the effect of a boundary from the interaction of the field with the medium, modeled as a collection of quantum harmonic oscillators, have been proposed in the literature \cite{HB92}.
On the contrary, in a truly microscopic approach (when dispersion interactions between the atoms/molecules of a macroscopic body are summed up), the presence of matter is described dynamically from a more fundamental point of view, in terms of its  Hamiltonian and the field-matter interaction Hamiltonian.

A microscopic pairwise summation (PWS) approach, consisting in summing up only the two-body components of the interaction, works correctly only for dilute dielectrics \cite{BMM99,Hamaker37,Marachevsky1}, but not in general for dense dielectrics and metals. The PWS approximation in fact does not take into account the many-body components of the dispersion interactions, or, equivalently, exchange of more than two virtual photons between the objects involved. The error made by the PWS approximation in some geometric configurations has been recently considered by comparison with exact macroscopic results \cite{BCAR13,Bennett14}.
A microscopic approach to Casimir interactions could be also relevant in order to understand discordant results in the literature about the attractive or repulsive character of the Casimir force or stresses for closed topologies such as a sphere \cite{attrrep,KK06}. In this case, however, the situations is made much more tricky by the absence of a uniquely defined minimum distance between the particles.

In this paper we develop a microscopic approach to Casimir and Casimir-Polder forces involving macroscopic metallic bodies, including two- and three-body components of the dispersion force (van der Waals/Casimir-Polder) between their constituents (metal nanoparticles). Specifically, we consider the two following configurations: metal nanoparticle/metal half-space and two metal half spaces. This will allow us to compare the role of two- and three-body components for such objects, and also gain some hints about the convergence rapidity of the many-body expansion for non-dilute systems such as a metal. We find that the two- and three-body contributions are of the same order of magnitude for metals; also, while the overall two-body contribution is attractive, the overall three-body contribution turns out to be repulsive.

This paper is organized as follows. In Sec. \ref{Sec:2} we outline our microscopic approach and derive the expressions of the  two- and three-body interaction between metal nanoparticles. These results will be then used in Sec. \ref{Sec:3} to evaluate from a microscopic point of view the two- and three-body contributions of dispersion interactions to the nanoparticle-wall Casimir-Polder interaction and to the wall-wall Casimir interaction, by summing up the dispersion interactions between their constituents (treated as metallic nanoparticles). Sec. \ref{Sec:4} will be devoted to our concluding remarks.

\section{The microscopic approach to Casimir-Polder and Casimir interactions}
\label{Sec:2}

In our microscopic approach we imagine to decompose a macroscopic metallic body into small metal spheres, or nanoparticles, of radius $\rho \,$; a typical value of their radius could be of the order of
some nanometer. We first evaluate their non-additive van der Waals/Casimir-Polder interaction in terms of their dynamical polarizability
$\alpha \left( \omega  \right)$. Finally, we sum these dispersion interactions over all the nanoparticles, maintaining $\mathcal{N}$-body components up to a given order.

In order to obtain the whole Casimir force from a microscopic theory, it is essential to consider non-additive effects too, so all $N$-body dispersion interactions should in principle be taken into account. The Casimir energy in the material is given by a $\mathcal{N}$-body expansion \cite{Buhmann2}:
\begin{equation}
\label{eq:1}
W=\sum\limits_{\mathcal{N} = 1}^\infty  {\frac{1}{{\mathcal{N}!}}\int\limits_V {{d^3}} {x_1}N\left( {{{\mathbf{x}}_1}} \right)...}
{\int\limits_V {{d^3}} {x_\mathcal{N}}N\left( {{{\mathbf{x}}_\mathcal{N}}} \right){U^{\left( \mathcal{N} \right)}}\left( {{{\mathbf{x}}_1},...,{{\mathbf{x}}_\mathcal{N}}} \right)} \, ,
\end{equation}
where $U^{\left(\mathcal{N} \right)} \left({{\bf{x}}_1,...,{\bf{x}}_\mathcal{N} } \right)$ is the $\mathcal{N}$-body dispersion interaction, and $N(\mathbf{x})$ the particle number density. We will now obtain the specific expressions of the two- and three-body components of the dispersion interaction between metal nanoparticles that will be used in the next Section.

We consider here the two- and three-body dispersion interactions between identical metal nanoparticles, both in the non-retarded and retarded limits.
Assuming these nanoparticles larger than the electron mean-free path, spatial dispersion can be neglected and their response to an electric field can be represented by a dielectric constant
$\varepsilon \left( \omega \right)$. We consider the metal as a free-electron gas,
where only the electric dipole moments due to free electrons contribute to the macroscopic polarization. According to the Drude-Sommerfeld model, we have
\begin{equation}
\label{eq:2}
\varepsilon \left( \omega  \right) = 1 - \frac{{\omega _p^2}}{{{\omega ^2} + i\omega \Gamma }} \, ,
\end{equation}
where ${\omega _p} = \sqrt {n{e^2}/{m_e}{\varepsilon _0}} $ is the plasma frequency of the metal, with $n$ the free electrons density, and  $\Gamma= v_F/l$ is a damping term, with $v_F$ the Fermi velocity and $l$ the electrons' mean-free path \cite{NH12}. The constant $\Gamma$ takes into account dissipative effects of the medium and therefore also its fluctuations.
The polarizability for these nanoparticles can be represented in the quasi-static limit (valid for a wavelength of light much greater than the characteristic size of the nanoparticle) as
\begin{equation}
\label{eq:3}
\alpha\left( \omega \right)=4\pi\epsilon_{0}\rho^{3}\frac{\epsilon\left( \omega \right)-1}{\epsilon\left( \omega \right)+2} \, ,
\end{equation}
where $\rho$ is the nanoparticle radius \cite{NH12,NC08}.
We consider the nanoparticles as rigid spheres without overlapping of their electron clouds (this condition also allows us to neglect spatial dispersion
in the dielectric constant).

The dispersion interaction between two identical isotropic polarizable particles in the vacuum in their ground state, at zero temperature, separated by a distance $r$, is \cite{CP48}
\begin{equation}
\label{eq:4}
 U^{(2)}\left( r \right) =  - \frac{\hbar }{{16\pi ^3 \varepsilon _0^2 }}\int\limits_0^\infty d\xi\frac{\alpha^{2}\left( i\xi \right)g_2\left( \xi r/c\ \right)}{r^6} \, ,
\end{equation}
where $g_2\left( x \right) = e^{ - 2x} \left( {3 + 6x + 5x^2  + 2x^3  + x^4 } \right)$ and $c$ is the speed of light.

\begin{figure}[bh]
\centering
\includegraphics[width=0.5\textwidth]{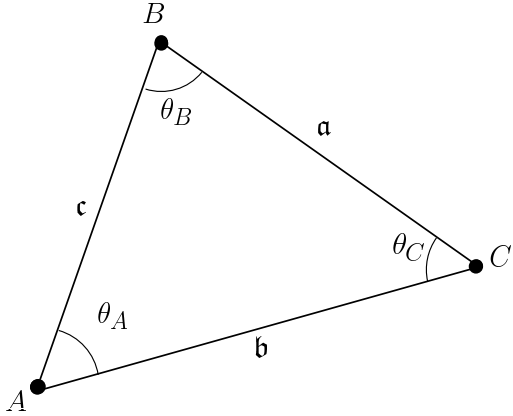}
\caption{Geometrical arrangement of the three metal nanoparticles.}
\label{Fig:1}
\end{figure}

Similarly, the three-body dispersion interaction between three identical isotropic polarizable particles in vacuum at zero temperature, is \cite{Buhmann2,AZ60,PT85}
\begin{equation}
U^{(3)}=\frac{\hbar}{64\pi^4\epsilon^{3}_{0}}\int^{\infty}_{0}d\xi\frac{\alpha^{3}\left(
i\xi \right)g_{3}\left( \mathfrak{a}\xi/c,\mathfrak{b}\xi/c,\mathfrak{c}\xi/c
\right)}{\mathfrak{a}^3\mathfrak{b}^3\mathfrak{c}^3}
\label{eq:5}
\end{equation}
where $\mathfrak{a}$, $\mathfrak{b}$, $\mathfrak{c}$ are the distances between the nanoparticles as shown in Fig. \ref{Fig:1}, and
\begin{eqnarray}g_3\left( {x,y,z} \right) &=& e^{-(x+y+z)}  \Big[ {3f\left( x \right)} f\left( y \right)f\left( z \right)
- g\left( x \right)f\left( y \right)f\left( z \right) \nonumber \\
&-& f\left( x \right)g\left( y \right)f\left( z \right)
- f\left( x \right)f\left( y \right)g\left( z \right)+ f\left( x \right)g\left( y \right)g\left( z \right){\cos ^2} \theta _A  \nonumber \\
&+& g\left( x \right)f\left( y \right)g\left( z \right)\cos ^2 \theta _B
+ g\left( x \right)g\left( y \right)f\left( z \right) \cos^2\theta _C \nonumber \\
&+& g\left( x \right)g\left( y \right)g\left( z \right)\cos \theta _A \cos \theta _B \cos \theta _C \Big] \, ,
\label{eq:6}
\end{eqnarray}
with $f(x)=1+x+x^2$ and $g(x)=3+3x+x^2$. $\theta _A$, $\theta _B$ and $\theta _C$ are the angles opposite to $\mathfrak{a}$, $\mathfrak{b}$, $\mathfrak{c}$, respectively, as shown in Fig. \ref{Fig:1}.

Simplified expressions can be obtained from Eq. (\ref{eq:5}) in two opposite regimes, where dispersion interactions reduce to simple power laws:
the non-retarded regime, when all distances between the particles are much smaller than the plasma wavelength  $\lambda_p = 2\pi c/\omega_p$ of the nanoparticles;
the retarded regime, if at least one interparticle distance is  much greater than $\lambda_p$.
In the retarded regime, many plasma oscillations occur during the time taken by the virtual photons involved in the interaction to travel between the nanoparticles, and this reduces the strength of the interaction.

We shall use these known expressions for the dispersion interactions in order to describe the interaction between our metallic nanoparticles. In the non-retarded limit (van der Waals), the $\xi$ integral in (\ref{eq:4}) and (\ref{eq:5}) restricts to a region where ${g_2}\left( {r\xi /c} \right) \simeq {g_2}\left( 0 \right)$, ${{g_3}\left( {\mathfrak{a}\xi /c,\mathfrak{b}\xi /c,\mathfrak{c}\xi /c} \right)}\simeq {g_3}\left( {0,0,0} \right)$, so the two- and three-body dispersion interactions take the simpler forms
\begin{eqnarray}
U_{non - ret}^{\left( 2 \right)} =  - \frac{{\sqrt 3 }}{4}\hbar \left( {{\omega _p} - \frac{{2\sqrt 3 \Gamma }}{\pi } + ...} \right){\rho ^6}\frac{1}{{{r^6}}},
\label{eq:7}
\end{eqnarray}
and
\begin{equation}
U_{non - ret}^{\left( 3 \right)} = \frac{{3\sqrt 3 }}{{16}}\hbar \left( {{\omega _p} - \frac{{8\sqrt 3 \Gamma }}{{3\pi }} + ...} \right){\rho ^9}
\frac{{1 + 3\cos {\theta _A}\cos {\theta _B}\cos {\theta _C}}}{{{\mathfrak{a}^3}{\mathfrak{b}^3}{\mathfrak{c}^3}}}.
\label{eq:8}
\end{equation}

These expressions are indeed Taylor expansions around $\Gamma /{\omega _p} = 0$, since for most metals $\Gamma  \ll {\omega _p}$ (for example, for gold $\omega _p  = 1.38 \cdot 10^{16} \, {\mbox{s}}^{-1}$ and $\Gamma  = 1.075 \cdot 10^{14}\, {\mbox{s}}^{-1}$).

In the retarded limit, the exponential factor in (\ref{eq:4}) and  (\ref{eq:5}) restricts the $\xi$-integral to a region where
the atomic polarizability can be approximated by its static value $\alpha \left( i\xi \right) \simeq \alpha \left( 0 \right)$. In this case, the two- and three-body dispersion interactions become
\begin{equation}
U_{ret}^{\left(2 \right)}  =  - \frac{{23\hbar c}}{{4\pi }}\rho ^6  \frac{1}{{r^7 }} \, ,
\label{eq:9}
\end{equation}

\begin{equation}
U_{ret}^{\left( 3 \right)} = \frac{{4\hbar c{\rho ^9}}}{\pi }f\left( {\mathfrak{a},\mathfrak{b},\mathfrak{c}} \right) \, ,
\label{eq:10}
\end{equation}
where $f\left( {\mathfrak{a},\mathfrak{b},\mathfrak{c}} \right)$ is a function depending only from the geometrical arrangement of the three nanoparticles
\begin{eqnarray}
f\left( {\mathfrak{a},\mathfrak{b},\mathfrak{c}} \right) &=& \frac{1}{{{\mathfrak{a}^3}{\mathfrak{b}^3}{\mathfrak{c}^3}\left( {\mathfrak{a} + \mathfrak{b} + \mathfrak{c}} \right)}}\Big[{{f_1}} +{f_2}\left( {\mathfrak{a},\mathfrak{b},\mathfrak{c}} \right){\cos ^2}\theta _A  \nonumber \\
&+& {f_2}\left( {\mathfrak{b},\mathfrak{c},\mathfrak{a}} \right){\cos ^2}\theta _B + {f_2}\left( {\mathfrak{c},\mathfrak{a},\mathfrak{b}} \right){{\cos }^2}\theta _C \nonumber \\
&+&   {f_3}\cos {\theta _A}\cos \theta _B \cos \theta _C \Big] \, ,
\label{eq:11}
\end{eqnarray}
with
\begin{eqnarray}
f_1 &=& 9 - 39\frac{{\sigma _2 }}{{\sigma _1^2 }} + 22\frac{{\sigma _3 }}{{\sigma _1^3 }} + 54\frac{{\sigma _2^2 }}{{\sigma _1^4 }} - 65\frac{{\sigma _2 \sigma _3 }}{{\sigma _1^5 }}
+ 20\frac{{\sigma _3^2 }}{{\sigma _1^6 }} \, , \nonumber \\
f_2 \left( {\mathfrak{a},\mathfrak{b},\mathfrak{c}} \right) &=& 3 \left[
\frac{{{\mathfrak{a}^2}}}{{\sigma _1^2}} + \frac{{3{\mathfrak{a}^2}\left( {\mathfrak{b} + \mathfrak{c}} \right)}}{{\sigma _1^3}} + \frac{{4\mathfrak{b}\mathfrak{c}\left( {3{\mathfrak{a}^2} - \mathfrak{b}\mathfrak{c}} \right)}}{{\sigma _1^4}}
-  \frac{{20\mathfrak{a}{\mathfrak{b}^2}{\mathfrak{c}^2}}}{{\sigma _1^5}} \right] \, , \nonumber \\
f_3 &=& 1 + 39\frac{{\sigma _2 }}{{\sigma _1^2 }} - 17\frac{{\sigma _3 }}{{\sigma _1^3 }} - 72\frac{{\sigma _2^2 }}{{\sigma _1^4 }} + 75\frac{{\sigma _2 \sigma _3 }}{{\sigma _1^5 }}
 - 20\frac{{\sigma _3^2 }}{\sigma _1^6 } \, .
 \label{eq:12}
\end{eqnarray}
and $\sigma _i  = \mathfrak{a}^i  + \mathfrak{b}^i  + \mathfrak{c}^i $.

The dependence on the geometrical parameters is the same as for dielectrics, well-known in the literature \cite{Buhmann2,AZ60,PT85}.
The three-body dispersion interaction can be attractive or repulsive. For example, for an equilateral triangular or a right triangle configuration, the non retarded force is repulsive, while the force is attractive when the three spheres are collinear. In general, the three-body contribution can be attractive or repulsive, depending on the specific geometry \cite{Buhmann2,PT85,Salam10}.

For perfect conductors ($\lambda_p \to 0$), dispersion interactions between nanoparticles are always in the retarded regime, regardless of their separation distance. This is the case we shall consider in the next Sections.

\section{Two- and three-body contributions of dispersion interactions to Casimir-Polder and Casimir forces}
\label{Sec:3}

We now consider explicitly the microscopic approach to Casimir-Polder and Casimir forces for two specific cases: a metallic nanoparticle near a perfectly conducting metallic half-space, and two metallic half spaces made of a perfect conductor. The results we obtain in the first case will be also the basis for the second one.

\subsection{Two- and three-body contribution to the Casimir-Polder energy between a metallic nanoparticle and a perfectly conducting half-space}

\begin{figure}[h]
\centering
\includegraphics[width=0.6\textwidth]{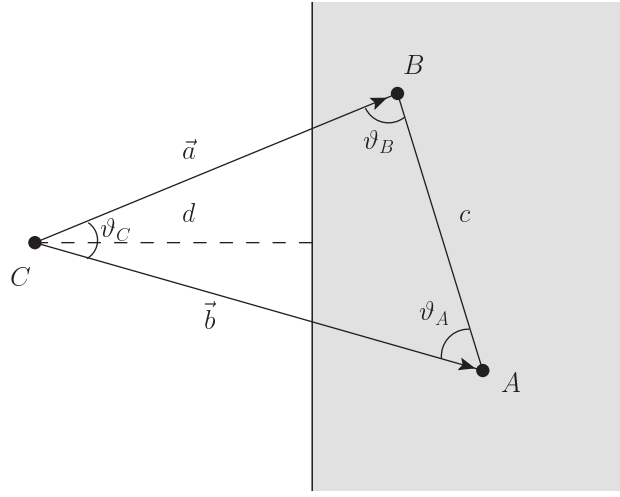}
\caption{\label{fig:2} Interaction between a nanoparticle $C$ and two generic nanoparticles in the metal half-space.}
\end{figure}

Let us first consider the Casimir-Polder interaction between a metallic nanoparticle $C$ at a distance $d$ from a perfectly conducting half-space, as shown in Fig. \ref{fig:2}.

We first analyze the two-body contribution to the Casimir-Polder force for this geometry, summing all two-body dispersion interactions between the metallic nanoparticle $C$ and a generic nanoparticle of the metallic half-space. We need to consider only the retarded interaction (\ref{eq:9}) because both the nanoparticle and the half-space are perfect conductors ($\lambda_p \to 0$). The coordinates are chosen such that the origin is kept fixed in $C$, and the z-axis is orthogonal to the half-space surface.
The two-body contribution to the nanoparticle-metallic half-space Casimir-Polder force is
\begin{equation}
W^{\left( 2 \right)}_{CP} = \int {U_{ret}^{\left( 2 \right)}NdV}  =  - \frac{69}{160\pi}\frac {\hbar c{\rho ^3}}{d^4} \, ,
\label{eq:13}
\end{equation}
where $N=\left( {4\pi {\rho ^3}/3} \right)^{-1}$ is the number of nanoparticles for unit volume and the integration over the volume has the following ranges: $0 \leqslant\theta < \pi/2$, $0 \leqslant \varphi \leqslant2 \pi$ and $r \geqslant d/\cos \theta$.

We now evaluate the three-body contribution to the Casimir-Polder interaction, by summing all three-body dispersion interactions between the nanoparticle $C$ and two generic nanoparticles in the metallic half-space (see Fig. \ref{fig:2}). Three-body interactions among three nanoparticles of the half space do not contribute to the interaction energy between the nanoparticle $C$ and the metal half-space (they only contribute to the half-space electromagnetic self-energy), and thus they will be not considered.
We use a spherical coordinate system with the position of two generic nanoparticles
of the half-space identified by their radial distance from $C$ and the polar and azimuthal angles in
\begin{eqnarray}&&\vec a = \left( {{\mathfrak{a}},{\theta _A},{\varphi _A}} \right) \nonumber \\
&&\vec b = \left( {{\mathfrak{b}},{\theta _B},{\varphi _B}} \right).
\end{eqnarray}
Here   $0 \leqslant {\theta _A} < \pi/2 $,  $\mathfrak{a}\geqslant d/{\cos \theta _A} $, $0 \leqslant {\varphi _A} \leqslant 2\pi $,
and analogous conditions hold for the nanoparticle B.
In this coordinate system, the distance $\mathfrak{c}$ between the nanoparticles $A$ and $B$ is given by
\begin{equation}
\mathfrak{c}=\sqrt{\mathfrak{a}^2 + \mathfrak{b}^2 - 2\mathfrak{a}\mathfrak{b}\left[ {\cos \theta _A \cos \theta _B + \sin \theta _A \sin \theta _B \cos \varphi } \right]} \, ,
\label{eq:14}
\end{equation}
where $\varphi= \varphi_A-\varphi_B$. The cosine of the angles $\vartheta_A,\vartheta_B,\vartheta_C$ can be obtained in terms of $\mathfrak{a},\mathfrak{b},\mathfrak{c}$,  using straightforward trigonometric relations.
The system is invariant under a generic rotation around the $z$-axis and thus
all geometrical parameters depend on the angles $\varphi_A$ and $\varphi_B$ only via
their difference $\varphi=\varphi_A-\varphi_B$.

Summing over the retarded three-body dispersion interactions (\ref{eq:10}), we obtain the three-body contribution
to the Casimir-Polder energy in the following form
\begin{equation}
\label{eq:15}
 W^{\left(3 \right)}_{CP}  = \frac{1}{{2!}}\int\limits_\Omega  {U_{ret}^{\left(3 \right)} N^2 dV_A dV_B}  = \frac{{2\hbar c}}{\pi }\rho ^9 N^2 K\left( {d,\lambda } \right)
\end{equation}
where $\rho$ is the nanoparticle radius, $N=\left( {4\pi {\rho ^3}/3} \right)^{-1}$ is the number of nanoparticles per unit volume and $K\left( {d,\lambda } \right)$ is the following function
\begin{eqnarray}
\label{eq:16}
&\ &K\left( {d,\lambda } \right)= \! \int\limits_0^{\pi /2} \! d\theta _A \int\limits_0^{ + \infty }\! d\mathfrak{a} \int\limits_0^{2\pi }\! d\varphi _A \int\limits_0^{\pi /2}\! d\theta _B \int\limits_0^{ + \infty } \! d\mathfrak{b} \int\limits_0^{2\pi }\! d{\varphi _B} \Theta \left( {\mathfrak{c} - \lambda } \right) \nonumber \\
&\times& \! \! \Theta \left( {\mathfrak{a} - d/\cos {\theta _A}} \right)\Theta \left( {\mathfrak{b} - d/\cos {\theta _B}} \right)
f\left( {\mathfrak{a},\mathfrak{b},\mathfrak{c}} \right){\mathfrak{a}^2}{\mathfrak{b}^2}\sin {\theta _A}\sin {\theta _B} \,.
\end{eqnarray}

In Eq. (\ref{eq:16}), $\Theta (x)$ is the Heaviside function, $\lambda$ the interparticle distance in the metal half-space and the function $f\left( {\mathfrak{a},\mathfrak{b},\mathfrak{c}} \right)$ has been defined in (\ref{eq:11}).
The first Heaviside function in the first line of Eq. (\ref{eq:16}) gives the condition $\mathfrak{c} \ge \lambda$, because the distance between two nanoparticles in the half-space must be larger than the interparticle distance; the two other Heaviside functions give the conditions $z_A \ge d$ and $z_B\ge d$ for the $z$ component of the position of particles A and B, respectively.

Although the integral in (\ref{eq:16}) is not defined for $\lambda=0$ (continuum case), we will consider its limit for
$\lambda \to 0$. We now show that the integrals giving $K\left( d \right) = \mathop {\lim }\limits_{\lambda  \to 0} K\left( {d,\lambda } \right)$ are finite in this limit apart from an additive constant that, being independent from the distance $d$ between the particle and the metal half space, does not contribute to the Casimir-Polder energy we are interested in. This also indicates that the divergence in the continuum limit is related to a Casimir self-energy related to the metallic half-space only, as it is expected from previous results concerning field divergences at boundaries for perfect conductors or field sources \cite{BP12,PRS13}.

In order to simplify the evaluation of the integral $K\left( d \right) $, we first take its derivative with respect to $d$, using the relation $\Theta '\left( x \right) = \delta \left( x \right)$. We obtain
\begin{eqnarray}
\label{eq:17}
\frac{{\partial K\left( d \right)}}{{\partial d}} &=&- 2\int\limits_0^{\pi /2} {d{\theta _A}} \int\limits_{d\sec {\theta _A}}^{ + \infty } {d\mathfrak{a}} \int\limits_0^{2\pi } {d{\varphi _A}} \int\limits_0^{\pi /2} {d{\theta _B}} \int\limits_0^{2\pi } {d{\varphi _B}} \nonumber \\
&\times& \left. \! \! \Theta \left( \mathfrak{c} \right) {f\left( {\mathfrak{a},\mathfrak{b},\mathfrak{c}} \right){\mathfrak{a}^2}{\mathfrak{b}^2}\sin \theta _A\tan \theta _B} \right|_{\mathfrak{b} = d/\cos \theta _B}\,\nonumber .\\
\end{eqnarray}
The factor 2 in (\ref{eq:17}) has been introduced because both cases $z_A=d$ and $z_B=d$ equally contribute.  With this procedure, the dimensionality of the integral has been reduced, fixing the
position of the nanoparticle $B$ on the plane $z_B=\mathfrak{b} \cos \theta_B = d$.
The integrand depends on the angles $\varphi_A$ and $\varphi_B$ only through
their difference $\varphi=\varphi_A-\varphi_B$, due to the symmetry of our system. With the substitutions $\varphi=\varphi_A-\varphi_B$ and $\varphi'=\varphi_A+\varphi_B$, we can use the relation
\begin{equation}
\int\limits_0^{2\pi } {d{\varphi _A}} \int\limits_0^{2\pi } {d{\varphi _B}} g\left( {{\varphi _A} - {\varphi _B}} \right) =   2\int\limits_0^{2\pi } {\varphi g\left( \varphi \right)} d\varphi
\, ,
\label{eq:18}
\end{equation}
where $g(\varphi )$ is a periodic function with period $2 \pi$. Using this expression in Eq. (\ref{eq:17})
and renaming $\varphi$ as $\varphi_A$, we obtain
\begin{eqnarray}
\frac{\partial K(d)}{\partial d}&=&- 4\int \limits_0^{\pi /2} \! d\theta _A \int \limits_{d/\cos \theta _A}^\infty \! d \mathfrak{a} \int\limits_0^{2\pi} \! d \varphi _A
\int\limits_0^{\pi /2} \! d\theta _B \Theta (\mathfrak{c})\nonumber \\
&\times& \! \! \left. f({\mathfrak{a},\mathfrak{b},\mathfrak{c}}) \mathfrak{a}^2 \mathfrak{b}^2 \varphi _A \sin \theta _A \tan \theta _B \right|_{\mathfrak{b} = d/\cos \theta_B,\varphi _B = 0} \nonumber \\
\, .
\label{eq:19}
\end{eqnarray}

We note that in the continuum limit $\lambda \to 0$, the function $\partial K(d)/ \partial d$ can depend only from the distance $d$, because there are no other distance scales involved, and must have the dimension of a
length to the power $-5$; thus it must have the following form
\begin{equation}
\label{eq:20}
\frac{\partial K(d)}{\partial d}= \frac \alpha {d^5} \, .
\end{equation}
The constant $\alpha$ can be evaluated numerically by a direct evaluation of the integral in
(\ref{eq:19}) for $d= 1$. We have evaluated it by a numerical integration with a symbolic algebra software using an adaptive procedure,
in order to deal properly with the singular behavior of the integrand in $\mathfrak{c}=0$.
We have obtained the following value for this constant:
$\alpha= -8.5 \pm 0.3$.
It follows that:
\begin{equation}
\label{eq:23}
K(d)=(2.1 \pm 0.1) \frac 1{d^4} \, .
\end{equation}
As mentioned, the (diverging) integration constant has not been considered because it yields a distance-independent self-energy, that does not contribute to the Casimir-Polder force.
The overall three-body contribution to the  Casimir-Polder energy is then
\begin{equation}
\label{eq:24}
W^{(3)}_{CP} \simeq (7.6 \pm 0.4) \cdot 10^{-2} \hbar c\rho^3\frac 1{d^4} \, .
\end{equation}

We can now compare (\ref{eq:24}) with (\ref{eq:13}), that is we are comparing the overall three- and two-body contributions to the atom-wall interaction energy.
We first note that, although the two-body contribution is always attractive (as shown by Eq. (\ref{eq:13})), the sign of (\ref{eq:24}) shows that the overall three-body contribution turns out to be repulsive and about one half of the two-body contribution. The distance dependence as $d^{-4}$ is the same for both contributions.

Our results (\ref{eq:13}) and  (\ref{eq:24}) give respectively  the overall two- and three-body contributions to the nanoparticle-wall Casimir-Polder force from a purely microscopic point of view, where the single components of the macroscopic metallic body are treated quantum-mechanically as field sources and not as a boundary condition. We can now compare them with known results extracted from a macroscopic approach \cite{Buhmann}.

The macroscopic retarded total Casimir energy between a metallic nanoparticle and a conductor half-space, described by a static dielectric function $\varepsilon$, is \cite{Buhmann}
\begin{eqnarray}
W &=& - \frac 3{16\pi }\int\limits_1^\infty  dv\left[ {\left( {\frac 2{v^2} - \frac 1{v^4}} \right)\frac{{\varepsilon v - \sqrt {\varepsilon  - 1 + v^2} }}{\varepsilon v + \sqrt {\varepsilon  - 1 + v^2} } + } \right. \nonumber \\
&\ &- \left. \frac 1 {v^4}\frac {v - \sqrt {\varepsilon  - 1 + {v^2}} }{v + \sqrt {\varepsilon  - 1 + {v^2}} } \right] \frac {\hbar c{\rho ^3}}{d^4} \, .
\label{eq:25}
\end{eqnarray}

Many-body contributions can be extracted from Eq. (\ref{eq:25}) by expanding the function
around the point $x= 4 \alpha N/\epsilon_0$ and expressing the relative dielectric function in terms of the quantity $\varepsilon=(3+2x)/(3-x)$, obtained from the Clausius-Mossotti formula. For our configuration, from the macroscopic theory we have the following expressions for the  two- and three-body contributions
\begin{equation}
W^{(2)}_{CP} =  - \frac {69}{160\pi} \frac {\hbar c{\rho ^3}}{d^4} \simeq -0.137  \frac {\hbar c \rho^3}{d^4} \, ,
\label{eq:26}
\end{equation}
\begin{equation}
W^{(3)}_{CP} = \frac {111}{448\pi } \frac {\hbar c{\rho ^3}}{d^4} \simeq 0.0789\frac {\hbar c\rho^3}{d^4} \, .
\label{eq:27}
\end{equation}

On the other hand, the total interaction energy obtained from (\ref{eq:25}) is
\begin{equation}
W_{CP} =  - \frac 3{8\pi} \frac {\hbar c\rho^3}{d^4} \simeq  - 0.119 \frac {\hbar c\rho^3}{d^4} \, .
\label{eq:28}
\end{equation}

Comparison of our results (\ref{eq:13}) and (\ref{eq:24}), obtained through our microscopic approach of summing the two- and three-body components of dispersion forces, with the macroscopic-approach results (\ref{eq:26}) and (\ref{eq:27}), clearly shows that
the two approaches are fully compatible, at least for the two- and three-body contributions to the Casimir-Polder energy.
However, the sum of two and three-body contributions is not sufficient to obtain the total Casimir-Polder force between the nanoparticle and the half-space given by (\ref{eq:28}), indicating a slow convergence of the many-body expansion (\ref{eq:1}) for metal bodies. Thus, we may conclude that higher-order non-additive components also play a significant role in determining the total interaction energy, and that the $\mathcal{N}$-body expansion (\ref{eq:1}) seems to converge quite slowly for metals, contrarily to the case of dilute dielectrics \cite{BMM99}.

In our evaluation we have included only electric dipole contributions to the dispersion interactions.
Eqs. (\ref{eq:13}) and (\ref{eq:24}) show that the summation of two- and three-body dipole dispersion interactions yields the same $d^{-4}$ distance dependence for their overall contribution to the nanoparticle/half-space interaction, and we expect the same should also hold for overall four-body and higher $\mathcal{N}$-body contributions.
On the other hand, higher multipole three-body interactions have been recently evaluated
\cite{TYSBM12,Salam13,Salam14}
and shown to decrease with the distance with a larger power law compared to three-body dipole interactions, both in the non-retarded and retarded regimes, for distances larger than the typical size of the (nano)particles involved, similarly to the two-body interactions case.
Also, the effect of summing up the $\mathcal{N}$-body interactions between the nanoparticles on the distance dependence of the overall interaction should be the same regardless of the multipole order. Thus, overall higher-multipole contributions should be much smaller than overall dipole contributions when the nanoparticle/half-space separation distance is larger than the typical size of the nanoparticles (a few nanometers), as we are assuming. For example, for such distances the overall three-body interaction involving quadrupole moments should be smaller than overall four-body dipole interaction, because the former decreases with distance with a higher power law.

We point out that our microscopic approach can be also extended to more complicated cases, for example different geometries. In the next Section we shall use the results obtained above in the case of two metallic half-spaces.

\subsection{Two- and three-body contribution to the Casimir force between two perfectly conducting metal half-spaces}

We now consider the relevant geometry of two half-spaces of a perfect conductor, separated by a distance $d$ along the $z$ direction. For this geometry, the Casimir energy for unit area $A$, evaluated using a macroscopic approach based on the zero-point field energy, is well known \cite{Casimir48}
\begin{equation}
\frac {W_{Cas}}A =  - \frac {\pi ^2}{720} \frac {\hbar c}{d^3} \simeq  - 0.0137 \frac {\hbar c}{d^3} \, .
\label{eq:29}
\end{equation}

It follows from (\ref{eq:29}) that the interaction of the two metal half-spaces with the zero-point electromagnetic field fluctuations leads to an
attractive force between them. This is also consistent with a recent general theorem stating that the Casimir force between two bodies related to each other by a reflection is always attractive \cite{KK06}.

We now want to consider the Casimir interaction for this geometry from a different point of view, evaluating the two- and three-body contributions of dispersion interactions between their metallic components to the Casimir energy, in analogy with the case discussed in the previous subsection. This microscopic approach will also clarify the importance of non-additive effects for this geometry, allowing us to compare their overall role with respect to the two-body components. The two-body contribution to the Casimir energy per unit area is straightforward
 \begin{eqnarray}
\frac {W^{(2)}_{Cas}}A &=& \int\limits_{ - \infty }^\infty  dx_1 \int\limits_{ - \infty }^\infty  dy_1 \int\limits_{ - \infty }^0 dz_1 \int\limits_d^\infty dz_2 N^2 U_{ret}^{(2)} \left( \left| \bf{r}_1 - \bf{r}_2 \right| \right) \nonumber \\
&=&  - \frac {69}{640 \pi^2}\frac {\hbar c}{d^3} \simeq  - 0.0109 \frac {\hbar c}{d^3} \, ,
\label{eq:30}
\end{eqnarray}
where subscripts $1$ and $2$ refer to coordinates relative to the two half spaces.
In the derivation of (\ref{eq:30}), we have integrated only the retarded two-body dispersion interactions, given by Eq. (\ref{eq:9}), because the two half-spaces are made of a perfect conductor and the interparticle distance is always larger than $d$. This result is known in the literature and, for example, it can be found in Ref. \cite{Milonni94}.

Comparison of (\ref{eq:30}) with (\ref{eq:29}) shows that pairwise summation accounts only for about $80\%$ of the total macroscopic Casimir force.
Thus pairwise summation does not hold for this geometry, and non-additive effects must play an essential role to ensure compatibility between macroscopic and microscopic models. It is therefore worth to consider in more detail, and to evaluate explicitly, the overall role of the many-body components.

\begin{figure}[h]
\centering
\includegraphics[width=0.5\textwidth]{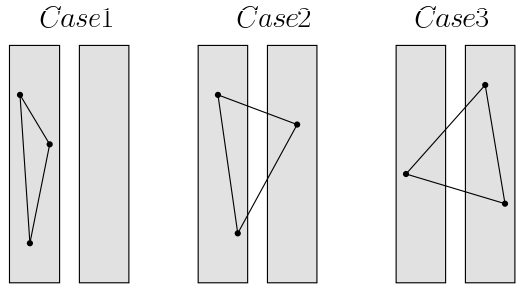}
\caption{\label{figslab} Different possible cases of a three-body interaction in the two metal half-spaces.}
\end{figure}

Next-order (non-additive) contribution to the Casimir energy is the three-body contribution of the dispersion interaction between all possible triplets of metal nanoparticles.  When we sum up retarded three-body dispersion interactions between three generic nanoparticles in the two conducting half-spaces, we must consider the three different cases shown in Fig. \ref{figslab}.\\
\\
Case 1 (and the similar one relative to the other half-space) gives a distance-independent contribution to the Casimir energy, and thus it does not contribute to the Casimir force between the two half spaces: hence, it will be neglected in our following evaluation. Cases 2 and 3 give equal contributions, and  we can exploit the result (\ref{eq:27}) obtained in the previous subsection for nanoparticle/half-space case. Using (\ref{eq:27}), after summation over all the nanoparticles $C$ of the other metallic half-space, we obtain
the three-body contribution to the Casimir energy per unit area
\begin{equation}
\frac {W^{(3)}_{Cas}}A = \frac {111}{896 \pi^2} \frac {\hbar c}{d^3} \simeq 0.0126 \frac {\hbar c}{d^3} \, .
\label{eq:31}
\end{equation}

Eq. (\ref{eq:31}) shows that the overall three-body contribution to the Casimir force is repulsive, contrarily to the overall two-body contribution which is attractive. Also, two- and three-body contributions are of the same order of magnitude giving, also in this case, a  strong indication of a slow convergence of the many-body expansion of the macroscopic Casimir energy. Therefore higher-order many-body contributions are expected to play an essential role too. Our result also indicates that both attractive and repulsive contributions play an essential role in this geometric configuration and that, for dense systems such as metal bodies, three-body and higher-order contributions have a role comparable to the usual two-body components. We expect this should be true also for other geometries.

\section{Conclusions}
\label{Sec:4}
In this paper we have developed a microscopic approach for Casimir and Casimir-Polder forces for metal macroscopic bodies, summing up the dispersion interactions between their constituents. In our approach, material bodies are treated in terms of the dynamical interactions between their constituents (metal nanoparticle), with fundamental dispersion interactions between them, and not as macroscopic objects giving only boundary conditions to the field operators. In particular, we have explicitly considered
two- and three-body dispersion interactions between the metal nanoparticles, both in the nonretarded and retarded regimes. Summing up these interactions, we have evaluated the overall two- and three-body contributions to the macroscopic Casimir-Polder and Casimir energy for two different configurations of metallic bodies: nanoparticle/half-space  and two half-spaces. Our expressions of the interaction for the nanoparticle/half-space geometry, obtained by the microscopic approach, are fully consistent with those that can be extracted from known macroscopic model results. We have found that the two-body interactions yield an attractive force, while the three-body interactions yield an overall repulsive force of comparable strength. This also suggests a quite slow convergence of the many-body expansion. Moreover, our results make clear the importance and role of three-body and higher dispersion interactions for the metallic systems considered, contrarily to the case of dilute dielectrics.

In the literature, to the best of our knowledge, the microscopic evaluation of the force has so far been developed only for dilute dielectrics or within the PWS approximation, where only two-body interactions are considered. Our model considers metals, in which the introduction of three-body (and higher many-body) interactions is necessary in order to obtain a thoroughly understanding of the physical problem and agreement with results extracted from macroscopic approaches, as our explicit results show. We wish to point out that our model can be also extended to other geometries or to non-dilute dielectrics, and clearly shows the importance of many-body dispersion interactions in the cases considered.

Our explicit evaluation shows that the pairwise approximation for the geometries we have considered is not valid and that non-additive effects must be taken into account. Specifically,  in the case of the two configurations considered we have also shown that, while the two-body dispersion interactions always lead to an attractive force, the overall three-body dispersion interaction leads to a repulsive one. Furthermore, overall three-body contributions are of the same magnitude as overall two-body contributions (at least in the cases here considered), indicating a slow convergence of the $\mathcal{N}$-body expansion.
Finally, we expect that the introduction of non-additive effects in the microscopic model could also clarify discordances in the literature concerning with the attractive or repulsive character of the Casimir force for some connected geometries, such as a perfectly conducting sphere \cite{attrrep,KK06}. We shall consider a microscopic approach to connected geometries, and discuss this important point, in a future publication.

\section*{Acknowledgments}

The authors wish to thank Stefan Buhmann and Lucia Rizzuto for many suggestions and encouragement on this work. Financial support by the Julian Schwinger Foundation, by MIUR and by CRRNSM is gratefully acknowledged.


\section*{References}

\begin{thebibliography}{99}
\bibitem{Casimir48} H.B.G. Casimir, Proc. K. Ned. Akad. Wet. \textbf{51}, 793 (1948).
\bibitem{Milonni94} P. W. Milonni, \emph{The Quantum Vacuum}, Academic Press, San Diego, 1994.
\bibitem{CP48} H. B. G. Casimir and D. Polder, Phys. Rev. \textbf{73}, 360 (1948).
\bibitem{Buhmann} S. Y. Buhmann, \emph{Dispersion forces I} (Springer, Heidelberg, 2013).
\bibitem{Buhmann2} S. Y. Buhmann, \emph{Dispersion Forces II} (Springer, Heidelberg, 2013).
\bibitem{CPP95} G. Compagno, R. Passante, and F. Persico, \emph{Atom-Field Interactions and Dressed Atoms} (Cambridge University Press, Cambridge 1995).
\bibitem{AT43} B. M. Axilrod and E.Teller, J. Chem. Phys. \textbf{11}, 299 (1943).
\bibitem{AZ60} M. R. Aub and S. Zienau, Proc. R. Soc. A \textbf{257}, 464 (1960).
\bibitem{PT85} E. A. Power and T. Thirunamachandran, Proc. R. Soc. A \textbf{401}, 267 (1985).
\bibitem{CP97} M. Cirone and R. Passante, J. Phys. B \textbf{30}, 5579 (1997).
\bibitem{RRE09} P. Rodriguez-Lopez, S. J. Rahi, and T. Emig, Phys. Rev. A \textbf{80}, 022519 (2009).
\bibitem{Salam10} A. Salam, \emph{Molecular Quantum Electrodynamics} (Wiley, Hoboken NJ, 2010).
\bibitem{MA14} R. Messina and M. Antezza, Phys. Rev. A {\bf 89}, 052104 (2014).
\bibitem{PP99} R. Passante and F. Persico, J. Phys. B \textbf{32}, 19 (1999).
\bibitem{SS12} K. V. Shajesh and M. Schaden, Int. J. Mod. Phys.: Conf. Series \textbf{14}, 521 (2012).
\bibitem{HB92} B. Huttner and S. M. Barnett, Phys. Rev. A \textbf{46}, 4306 (1992).
\bibitem{BMM99} I. Brevik, V. N. Marachevsky, and K. A. Milton, Phys. Rev. Lett. \textbf{82}, 3948 (1999).
\bibitem{Hamaker37} H. C. Hamaker, Physica \textbf{4}, 1058 (1937).
\bibitem{Marachevsky1}V. N. Marachevsky, Phys. Scr. \textbf{64}, 205 (2001).
\bibitem{BCAR13} A. F. Bitbol, A. Canaguier-Durand, A. Lambrecht, and S. Reynaud, Phys. Rev. B \textbf{87}, 045413 (2013).
\bibitem{Bennett14} R. Bennett, Phys. Rev. A \textbf{89}, 062512 (2014).
\bibitem{attrrep} T. H. Boyer, Phys. Rev. \textbf{174}, 1764 (1968);
K. A. Milton, L. L. DeRaad Jr., and J. Schwinger, Ann. Phys. (N.Y.) \textbf{115}, 388 (1978);
M. P. Hertzberg, R. L. Jaffe, M. Kardar, and A. Scardicchio, Phys. Rev. Lett. \textbf{95}, 250402 (2005).
\bibitem{KK06} O. Kenneth and I. Klich, Phys. Rev. Lett. \textbf{97}, 160401 (2006).
\bibitem{NH12} L. Novotny and B. Hecht, \emph{Principles of Nano-Optics}, (Cambridge University Press, Cambridge 2012).
\bibitem{NC08} L. Novotny and C. Henkel, Optics Letters, \textbf{33}, 1029 (2008).
\bibitem{BP12} N. Bartolo and R. Passante, Phys. Rev. A \textbf{86}, 012122 (2012).
\bibitem{PRS13} R. Passante, L. Rizzuto, and S. Spagnolo, Eur. Phys. J. C \textbf{73}, 2419 (2013).
\bibitem{TYSBM12} L. Y. Tang, Z. C. Yan, T. Y. Shi, J. F. Babb, and J. Mitroy, J. Chem. Phys. \textbf{136}, 104104 (2012).
\bibitem{Salam13} A. Salam, J. Chem. Phys. \textbf{139}, 244105 (2013).
\bibitem{Salam14} A. Salam, J. Chem. Phys. \textbf{140}, 044111 (2014).
\end{thebibliography}
\end{document}